\let\includefigures=\iftrue
%
\let\includefigures=\iffalse
%
\let\useblackboard=\iftrue
%
%
%
\input harvmac.tex
\message{If you do not have epsf.tex (to include figures),}
\message{change the option at the top of the tex file.}
\input epsf
\epsfverbosetrue
\def\fig#1#2{\topinsert\epsffile{#1}\noindent{#2}\endinsert}
\def\fig#1#2{}
%
\def\Title#1#2{\rightline{#1}
\ifx\answ\bigans\nopagenumbers\pageno0\vskip1in%
\baselineskip 15pt plus 1pt minus 1pt
\else
\def\listrefs{\footatend\vskip 1in\immediate\closeout\rfile\writestoppt
\baselineskip=14pt\centerline{{\bf References}}\bigskip{\frenchspacing%
\parindent=20pt\escapechar=` \input
refs.tmp\vfill\eject}\nonfrenchspacing}
\pageno1\vskip.8in\fi \centerline{\titlefont #2}\vskip .5in}

\ifx\answ\bigans\def\tcbreak#1{}\else\def\tcbreak#1{\cr&{#1}}\fi
\useblackboard
\message{If you do not have msbm (blackboard bold) fonts,}
\message{change the option at the top of the tex file.}
\font\blackboard=msbm10 scaled \magstep1
\font\blackboards=msbm7
\font\blackboardss=msbm5
\newfam\black
\textfont\black=\blackboard
\scriptfont\black=\blackboards
\scriptscriptfont\black=\blackboardss
\def\Bbb#1{{\fam\black\relax#1}}
\else
\def\Bbb{\bf}
\fi
%
\def\yboxit#1#2{\vbox{\hrule height #1 \hbox{\vrule width #1
\vbox{#2}\vrule width #1 }\hrule height #1 }}
\def\fillbox#1{\hbox to #1{\vbox to #1{\vfil}\hfil}}
\def\ybox{{\lower 1.3pt \yboxit{0.4pt}{\fillbox{8pt}}\hskip-0.2pt}}
\def\comments#1{}

\def\BZ{\Bbb{Z}}
\def\p{\partial}

\def\tr{{\rm tr\ }}

\Title{\vbox{\baselineskip12pt
\hfill{\vbox{
\hbox{BROWN-HET-1012\hfil}
\hbox{hep-th/9606091}}}}}
{\vbox{\centerline{Gluino Condensation in F-theory}
\vskip20pt
\centerline{}}}
\centerline{Miao Li}
\smallskip
\centerline{Department of Physics}
\centerline{Brown University}
\centerline{Providence, RI 02912}
\centerline{\tt li@het.brown.edu}
\bigskip
\noindent
Coupling of bilinears of gauginos on D-branes to anti-symmetric tensor
fields can be deduced by comparing F-theory compactified to 8 dimensions
with the heterotic string on $R^8\times T^2$. Application to SUSY breaking
in F-theory triggered by gaugino condensation is discussed. Compactification
to four dimensions with a warp factor is briefly discussed. In this case
some 3-branes whose world-volume coincides with the open spacetime must 
be introduced, in addition to 7-branes.

\Date{June 1996}
\nref\cumrun{C. Vafa, ``Evidence for F-theory,'' hep-th/9602022.}
\nref\mv{D. Morrison and C. Vafa, ``Compactifications of F-theory
on Calabi-Yau Threefolds-I,II,'' hep-th/9602114, hep-th/9603161.}
\nref\bers{M. Bershadsky, K. Intriligator, S.~ Kachru,
D.~R.~Morrison, V.~Sadov and C.~Vafa, ``Geometric Singularities
and Enhanced Gauge Symmetries,'' hep-th/9605200.}
\nref\km{D. Kutasov and E. Martinec, `` New Principles for String/Membrane 
Unification,'' hep-th/9602049; D.~Kutasov, E.~Martinec and M. 
~O'Loughlin, ``Vacua of M-theory and N=2 strings,'' hep-th/9603116.} 
\nref\sw{N. Seiberg and E. Witten, ``Comments on String Dynamics 
in Six Dimensions,'' hep-th/9603003.}
\nref\ew{E. Witten, `` Phase Transitions In M-Theory And F-Theory,''
hep-th/9603150.}
\nref\witten{E. Witten, ``Non-perturbative Superpotentials In
String Theory,'' hep-th/9604030.}
\nref\sadov{V. Sadov, `` Generalized Green-Schwarz mechanism in F theory,''
hep-th/9606008.} 
\nref\dthree{ A. A. Tseytlin, ``Self-duality of Born-Infeld action and 
Dirichlet 3-brane of type IIB superstring theory,'' hep-th/9602064;
D.~P.~Jatkar and S.~K.~Rama, ``F-theory from Dirichlet 3-branes,''
hep-th/9606009.} 
\nref\dl{M. R. Douglas and M. Li, to appear.}
\nref\drsw{M. Dine, R. Rohm, N. Seiberg and E. Witten, Phys. Lett.
B156 (1985) 55; J.~P.~ Derendinger, L.~E.~Ibanez and  H.~P.~Nilles,
Phys. Lett. B155 (1985) 65.}
\nref\mdl{M. R. Douglas and M. Li, ``D-brane Realization of ${\cal N}
=2$ Super Yang-Mills Theory in Four Dimensions,'' hep-th/9604041.} 
\nref\ss{A. Salam and E. Sezgin, Phys. Lett. B154 (1985) 37; M.~Awada
and P.~K.~Townsend, Phys. Lett. B156 (1985) 51.}
\nref\john{J. H. Schwarz, Nucl. Phys. B226 (1983) 269.}
\nref\greens{B. Greene, A. Shapere, C. Vafa and S.-T. Yau,
Nucl. Phys. B337 (1990) 1.}
\nref\ggp{G. Gibbons, M.B. Green and M.J. Perry, hep-th/9511080.}
\nref\sen{A. Sen, ``F-theory and Orientifolds,'' hep-th/9605150;
T.~Banks, M.~R.~Douglas and N.~Seiberg, ``Probing F-theory With Branes,''
hep-th/9605199.}
\nref\dasm{K. Dasgupta and S. Mukhi, ``F-theory at Constant Coupling,''
hep-th/9606044.} 
\nref\bsv{M. Bershadsky, V. Sadov and C. Vafa, ``D-Branes and Topological
Field Theories,'' hep-th/9511222.}
\nref\bbs{K. Becker, M. Becker and A. Strominger, ``Fivebranes, Membranes 
and Non-Perturbative String Theory,'' hep-th/9507158.}
\nref\andy{A. Strominger, Nucl.Phys. B274 (1986) 253.}
\nref\bb{ K. Becker, M. Becker, ``M-theory on Eight-manifolds,''
hep-th/9605053.}
\nref\nati{N. Seiberg, `` IR Dynamics on Branes and Space-Time Geometry,''
hep-th/9606017.}
Vafa's F-theory \cumrun\ provides an elegant way to do string 
compactifications \mv\ \bers. Geometrically, a compactification
is realized by an elliptically fibered Calabi-Yau space. The sum of dimension
of the base space  and that of the noncompact spacetime is 10.
Together with the hidden torus, the total dimension is 12. 
When the torus is degenerate, one finds a collection of coincident
7-branes. It is possible that the theory can be formulated in
12 dimensions \cumrun\ \km\ \dthree. For discussions on other aspects of the
theory, see \sw\ \ew\ \witten\ \sen\ \sadov\ \dasm.
In its original formulation, the compact space is
an elliptically fibered K3 with base space $S^2$ \cumrun. It was conjectured
to be dual of the heterotic string compactified on $R^8\times T^2$.
On the F-theory side, 24 D7-branes are required. At a generic point
of the moduli space, all 24 7-branes are separated. Naively,
one expects a gauge group $U(1)^{24}$, contradicting the fact that
the rank of the gauge group of 8 dimensional heterotic string
is 20. Indeed, some linear combinations of abelian gauge fields
on 7-branes are Higgsized \dl, as a result of couplings between
closed string fields and D-brane fields. In this paper, we point out
that as a result of duality between F-theory and heterotic string,
some gaugino bilinears are coupled to R-R anti-symmetric tensor fields.
These couplings can be computed in principle as mixed amplitudes
of brane fields and closed string fields on a disk. By T-duality,
one deduces similar couplings on D-branes other than 7-branes.
The most interesting application of these couplings is supersymmetry
breaking triggered by gaugino condensation, much as what happens
in heterotic string in 4 dimensions \drsw.

We start with a comparison between the low energy effective action
of 8 dimensional heterotic string and IIb string compactified on
$S^2$. Let the area of this $S^2$ measured in 10D Einstein metric
be $V(S^2)$. The relation between the 8D heterotic string coupling 
constant $\lambda_8$ and $V(S^2)$ is $\lambda_8=V(S^2)$, as pointed out 
in \mdl\ where no derivation was given (see also \witten). Here 
we give a simple derivation of this relation.
Let $G_8$ denote the 8D Einstein metric, and $G_{10}$ the
10D IIb Einstein metric. By comparing the 10D Einstein action to
the 8D Einstein action, one deduces
$$G_{10}=[V(S^2)]^{-1/3}G_8.$$
It is easy to see that the gauge field action on the heterotic side,
written in terms of 8D Einstein metric, is
\eqn\comp{\int d^8x\sqrt{g_8}\lambda_8^{-2/3}F^2.}
Interpreted as gauge field on a 7-brane, the action is
$$\int d^8x\sqrt{g_s}\lambda^{-1}F^2=\int d^8x\sqrt{g_{10}}F^2,$$
where $G_s$ is the 10D IIb string metric, and $\lambda$ the 10D
IIb string coupling constant. On the L.H.S. of the above equation,
$F^2$ is constructed using $G_s$, and the same quantity on the
R.H.S. is constructed using $G_{10}$. Finally, using $G_{10}=[V(S^2)]^{-1/3}
G_8$, one finds 
$$\int d^8x\sqrt{g_{10}}F^2=\int d^8x\sqrt{g_8}[V(S^2)]^{-2/3}F^2.$$
This compared with \comp\ tells us that $\lambda_8=V(S^2)$. So 
the relation between the 10D IIb Einstein metric and 8D Einstein 
metric can be written as 
\eqn\rela{G_{10}=\lambda_8^{-1/3}G_8.}
This relation will be relevant in our derivation of coupling of
gaugino bilinears on a 7-brane to antisymmetric tensor fields.

Now we are ready to derive the coupling of the bilinears of gauginos
on a 7-brane and R-R tensor fields. First, let $\chi_8$ be a
gaugino in the heterotic string with standard kinetic term
$1/2\int d^8x\sqrt{g_8}\bar{\chi}_8\gamma^\mu\p_\mu\chi_8$. The
low energy action of 8D ${\cal N}=1$ supergravity coupled to vector
multiplets was discussed in \ss. In particular, there is a coupling
between $\bar{\chi}_8\gamma^{\mu\nu\rho}\chi_8$ and $H_{\mu\nu\rho}$,
the strength of the standard antisymmetric tensor field:
\eqn\coup{-{1\over 24}\int d^8x\sqrt{g_8}\lambda_8^{-2/3}H_{\mu\nu\rho}
\bar{\chi}_8\gamma^{\mu\nu\rho}\chi_8,}
where we have used the fact that the scalar field $\sigma$ in the 
first reference of \ss\ has a relation to $\lambda_8$ as $e^\sigma
=\lambda_8^{-2/3}$.

On the F-theory side, there is no zero mode from $B_{\mu\nu}$ and
$C^{(2)}_{\mu\nu}$, as a result of the nontrivial $SL(2,\BZ)$
bundle over $S^2$ \cumrun\ \dl. However, in the IIb theory there
is a rank 4 R-R tensor field $C^{(4)}$. Since it is $SL(2,\BZ)$ singlet,
it has a zero mode. When restricted to $R^8$, one obtains a rank
4 tensor field, still denoted by $C^{(4)}$, and a rank 2 tensor
field which is dual to $C^{(4)}$ in 8 dimensions. Apparently, it is the 
dual of $C^{(4)}$ that corresponds to $B$, the standard rank 2 tensor
field in the heterotic string. If the conjectured duality holds,
then the coupling \coup\ tells us that there ought to be a coupling
between the field strength of $C^{(4)}$ and the bilinear of the gaugino.
Naturally, we would like to cast the coupling in terms of the original
10D IIb Einstein frame. Let $\chi$ be the gaugino on a 7-brane, normalized
against the IIb Einstein frame. Let $\chi^{(3)}=\bar{\chi}\gamma_{\mu
\nu\rho}\chi dx^\mu\wedge dx^\nu\wedge dx^\rho$, where the gamma matrices
are also defined in term of the IIb Einstein frame. We shall
now show that there is the coupling
\eqn\coupd{-{1\over 24\times 5!}\int H^{(5)}\wedge\chi^{(3)},}
on the 7-brane. 

All we have to do is to obtain the relation between
$H$ and $H^{(5)}$ with the correct normalization. To do this,
recall from \ss\ that the Einstein equation in 8D when $H$ is not zero
can be written schematically as
\eqn\compa{R_{\mu\nu}=\lambda_8^{-4/3}\left(g_{\mu\nu}H^2+H_{\rho\lambda\mu}
H^{\rho\lambda}_{\quad\nu}\right),}
where everything is written in terms of the 8D Einstein metric.
The Einstein equation when $H^{(5)}$ nonvanishing was written down in \john.
Also schematically, it is
$$R_{\mu\nu}=H^{(5)}_{\mu\mu_1\dots\mu_4}H_\nu^{(5)\mu_1\dots\mu_4},$$
where everything is written in terms of the 10D IIb Einstein metric.
Given the relation \rela, this is
$$R_{\mu\nu}=\lambda_8^{4/3}H^{(5)}_{\mu\mu_1\dots\mu_4}
H_\nu^{(5)\mu_1\dots\mu_4}$$
when written in 8 dimensions using the 8D Einstein metric. Note that
the Ricci tensor $R_{\mu\nu}$ is scale invariant. Compared with
\compa, the above equation indicates that the correction relation between
$H^{(5)}$ and $H$ is
\eqn\corrr{H^{(5)}=\lambda_8^{-4/3}\quad ^*H,}
where the dual is defined against the 8D Einstein metric. To finally
derive \coupd, we need the relation between $\chi_8$ and $\chi$. Using
\rela\ it is easy to see that $\chi_8=\lambda_8^{-7/12}\chi$. Substituting
this relation and \corrr\ into \coup\ one derives \coupd. Despite
all strange powers of $\lambda_8$, the final result is independent of
this 8D heterotic string coupling constant.

The coupling \coupd, derived from duality between F-theory and 8D
heterotic string, can be computed by the standard string technique.
In order to see that this is indeed possible, we want to recast the
coupling in terms of the 10D IIb string metric. $H^{(5)}$ remains
the same. Again let $\lambda$ be the IIb string coupling constant in
10 dimensions. The relation between the 10D Einstein metric and
the 10D IIb string metric, $G_s$, is $G_{10}=\lambda^{-1/2}G_s$. And
the relation between $\chi$ and $\chi_s$, normalized against the
string metric, is $\chi=\lambda^{7/8}\chi_s$. Let $\chi^{(3)}_s$
be the corresponding bilinear defined in string metric. So
there is relation $\chi^{(3)}=\lambda\chi^{(3)}_s$. Thus, the coupling
\coupd\ becomes
\eqn\coups{-{1\over 24\times 5!}\int \lambda H^{(5)}\wedge\chi_s^{(3)}.}
The weight $\lambda$ is expected, as the corresponding amplitude
is a disk amplitude with insertion of a R-R closed string vertex
operator and two open string vertex operators on the boundary.
(For a standard propagator, each fermion vertex operator carries
a weight $\sqrt{\lambda}$, the R-R vertex operator carries a 
weight $\lambda$, and the disk contributes a weight $\lambda^{-1}$.)
It is gratifying that the coupling, which we have derived via a 
convoluted route of argument, agrees with standard string
calculation. This certainly lends another piece of evidence
for duality between F-theory and heterotic string.

Before turning to discussion of other similar couplings, it may
be useful to point out that, according to \john, $H^{(5)}$ is
not simply $5dC^{(4)}$. Since under gauge transformation, say,
$\delta C^{(2)}=d\epsilon$, $C^{(4)}$ is not invariant. The
correct definition of $H^{(5)}$ is the following \john
\eqn\fstr{H^{(5)}=5\left(dC^{(4)}-{3\kappa\over 4}(B\wedge
dC^{(2)}-C^{(2)}\wedge dB)\right),}
where $B$ is the NS-NS antisymmetric tensor field. Apparently,
this definition is invariant under $SL(2,\BZ)$. $\kappa^2$ is
the 10D Newton constant as defined in \john. We have also changed
notations of \john\ to conform with more conventional notations.

On a 7-brane, naturally one wonders whether couplings $\int H^{(3)}
\wedge \chi^{(5)}$ and $\int H^{(7)}\wedge \chi^{(1)}$ also exist.
The following simple argument quickly rules out the first coupling.
For a 7-brane defined as a D-brane on which a $(1,0)$-string ends,
$C^{(2)}$ is not invariant under the monodromy created by presence 
of the 7-brane. After following the loop encircling the 7-brane
in the transverse space, $C^{(2)}\rightarrow C^{(2)}-B$, so
$H^{(3)}\rightarrow H^{(3)}-H$ and the first coupling is not defined
canonically. This argument does not rule out the second coupling.
$H^{(7)}$ is the field strength of $C^{(6)}$ which is dual to
$C^{(2)}$. Under the monodromy, $C^{(6)}$ is invariant while
$B^{(6)}$ is not. Note that the coupling $\int H^{(7)}\wedge \chi^{(1)}$
is not forbidden by duality between F-theory and the heterotic string
in 8 dimensions. When restricted on the world-volume of the
7-brane, $H^{(7)}$ can be written as dual of $\p_\mu C^{(2)}_{ij}$.
Now $C^{(2)}_{ij}$ does not have a zero mode on $S^2$, thus this
coupling does not give rise to a coupling to a massless bosonic
field in 8 dimensions. Despite all this, we believe that this
coupling does not exist. A simple argument is that when written against
the 10D Einstein metric, one would obtain a strange weight
$\sqrt{\lambda}$ for the coupling. An explicit calculation is
necessary to assure this claim.

Similar couplings on other D$p$-branes can be deduced from \coupd\
using T-duality. Consider, for example, a 6-brane in the IIa theory.
A 6-brane can be obtained by wrapping a 7-brane on a circle $S^1$
and T-dualizing on this circle. Coupling \coupd\
reduces to the couplings
\eqn\dsix{\int H^{(4)}\wedge\chi^{(3)}+\int H^{(5)}\wedge\chi^{(2)},}
where $H^{(5)}$ is simply the original $H^{(5)}$ restricted in
noncompact spacetime $R^9$, and $H^{(4)}$ is its dual in 9 dimensions.
Interpreted in the IIa theory, $H^{(4)}$ is the field strength
of $C^{(3)}$. The first coupling in \dsix\ is lifted to 10 noncompact
spacetime as 
\eqn\fsix{\int  H^{(4)}\wedge\chi^{(3)},}
while the second coupling, when lifted to 10 noncompact dimensions, is
\eqn\ssix{\int d^7x H^{(4)}_{\mu\nu i j}\bar{\chi}\gamma_k\gamma_{\mu\nu}
\chi\epsilon^{ijk},}
where $i,j,k$ are indices tangent to the 3 transverse dimensions.
Similar couplings on lower dimensional D-branes can be obtained 
in a similar way.

We now turn to a discussion of gluino condensation. When compactified
on an elliptic Calabi-Yau space, enhanced gauge symmetry is obtained
when several 7-branes coincide. Exactly what gauge group occurs
is determined by the type of singularity \mv\ \bers. The coupling
\coupd\ certainly generalizes to
\eqn\coupn{-{1\over 24\times 5!}\int H^{(5)}\wedge\tr\chi^{(3)},}
for a non-abelian gauge group. Again this is similar to what happens
in the heterotic string. Now, one might think that further discussion
of gluino condensation and SUSY breaking triggered is redundant,
since it may be exactly parallel to that of \drsw. This is not the
case in F-theory. The reason is simply that the original discussion
of \drsw\ depends on Calabi-Yau threefold compactification. While
to obtain a ${\cal N}=1$ 4D string theory, the F-theory is going
to be compactified on a Calabi-Yau fourfold. The base space of
this fourfold has 3 complex dimensions, but is not a Calabi-Yau
threefold itself. Furthermore, the singular locus which 7-branes
wrap around in general has different geometric features than those
of a Calabi-Yau threefold. 

When F-theory is compactified on a Calabi-Yau $(n+1)$-fold, the unbroken
supersymmetry is believed to be ${\cal N}=1$ supersymmetry in 
$10-2n$ dimensions. Although there is no doubt that this must be true,
local geometric conditions for unbroken SUSY has not been stated
explicitly in the existing literature.
Consider the simplest case when the Calabi-Yau is $K3$. This is
a IIb theory compactified on $S^2$. Existence of 7-branes 
induces a metric \greens\ as well as an auxiliary U(1) gauge
field \john\ on $S^2$. The supersymmetry parameter in the IIb theory
has a charge $1/2$ with respect to the U(1) gauge field. For a 7-brane, 
the spin connection is
exactly equal to the auxiliary U(1) connection \ggp. So
for a 2d chiral spinor, there exists a covariantly constant spinor.
This spinor is actually constant and gives rise to the 8D
${\cal N}=1$ supersymmetry. For a general Calabi-Yau  $(n+1)$-fold,
the base space $B$ is a dimension $n$ complex manifold. Now the
field strength of the auxiliary $U(1)$ gauge field has only 
nonvanishing components $F_{i\bar{j}}$, where we use complex 
coordinates $z^i$ on the base space. This is easily seen using
definition of the $U(1)$ gauge field in \john\ and the fact that
the complex coupling $\tau$ is a holomorphic function on the
base space. Existence of a single covariantly constant spinor on the
base space then implies that $R_{i\bar{j}}\epsilon =F_{i\bar{j}}
\epsilon$ and $R_{ij}\epsilon =R_{\bar{i}\bar{j}}\epsilon =0$.
The second condition is met when the base space is K\"ahler.
For a K\"ahler manifold, the holonomy group is $U(n)$.
The first condition implies that the $n\times n$ Hermitian
matrix $R_{i\bar{j}}$ has exactly one eigenvalue equal
to $F_{i\bar{j}}$. Note that we have assumed that $\epsilon$
has a positive chirality, and so is in the fundamental representation
of the holonomy group $U(n)$. 

In addition to the above condition, there is a further condition for
the unbroken SUSY. This comes from SUSY transformation of the dilatino
\john: 
$$\delta\lambda =\gamma^Mp_M\epsilon^*+\dots$$
where the dots denote terms associated to field strengths of antisymmetric
tensor fields $B$ and $C^{(2)}$. These are vanishing upon 
compactification; $p_M$ is a current constructed from $\tau$. For our
purpose, it is enough to know that $p_M\sim \p_M\tau$. Since $\tau$
is holomorphic, so $\delta\lambda\sim \gamma^ip_i\epsilon^*$, where
$\gamma^i$ are gamma matrices in holomorphic tangent directions. If
$\gamma^i\epsilon^*=0$, or $\gamma^{\bar{i}}\epsilon=0$, then $\delta
\lambda =0$ is satisfied. Using this
we find that $R_{i\bar{j}}\epsilon=R_{i\bar{j}}^{\quad l\bar{k}}g_{l\bar{k}}
\epsilon$. Then the condition derived in the previous paragraph
says that the Ricci tensor of the base space is equal to $F_{i\bar{j}}$.
This is not very strong condition, given that $\gamma^{\bar{i}}\epsilon
=0$. Concrete cases where $\tau$ is a constant have been discussed
recently \sen\ \dasm, where it is find that the base space
is always a flat orbifold of torus. This certainly conforms with
the condition presented above.
The condition $\gamma^{\bar{i}}\epsilon=\gamma_i\epsilon=0$
can be rephrased as stating that $\epsilon$ is the ground state, if
$\gamma_i$ are regarded as annihilation operators and $\gamma_{\bar{i}}$
as creation operators. 

There may be solutions other than $\gamma_i\epsilon=0$, this can be
seen by observing $\delta\lambda=p_i\gamma^i\epsilon^*$ using
a Fock basis for $\epsilon^*$.
However, when the Ricci tensor is equal to $F_{i\bar{j}}$, ours
is certainly a solution, and there must be no other solutions, otherwise
the unbroken SUSY is larger than ${\cal N}=1$. For the rest of
this paper, we shall assume $R_{i\bar{j}}=F_{i\bar{j}}$. 
We do not know whether our condition is sufficiently generic.
Further exploration of this issue will involve detailed study
of elliptic fibrations.

In the rest of this paper, we assume a compactification on a Calabi-Yau 
fourfold in which enhanced gauge symmetry occurs. Mathematically, 
there is a four dimensional
sub-manifold of the base space on which the fiber tori degenerate.
Physically, this sub-manifold is part of 7-brane world-volume. The
effective field theory in the remaining four noncompact dimensions
is a ${\cal N}=1$ Super Yang-Mills theory. Typically, gluino 
condensation will occur, and we want to know whether SUSY is broken
as a result of the condensation. The simplest signal is the appearance
of a  goldstino. The original dilatino can not be part of the 
goldstino. This is because, on the one hand, it is most 
likely that zero
modes of dilatino are paired with some linear combinations of
gauginos and gain mass, as what actually happens in 8 dimensions \dl.
On the other hand, the SUSY transformation $\delta\lambda$, as
mentioned before, contains terms depending on field strengths of
rank 2 anti-symmetric tensor fields. The supercovariant version
of these field strengths contain only bilinear of gravitino, and
bilinear mixing gravitino and dilatino \john. Aside from gauginos themselves,
one candidate for goldstino is the new dilatino which must
be a linear combination of gauginos. Since the SUSY transformation
of this dilatino has not been worked out, we will take a look at 
the gravitino, whose  SUSY transformation is \john
\eqn\gravi{\delta\psi_M=D_\mu\epsilon +{i\over 480}\gamma^{M_1\dots
M_5}\gamma_M\epsilon\hat{H}^{(5)}_{M_1\dots M_5}+\dots,}
where $\hat{H}^{(5)}$ is the supercovariant version of $H^{(5)}$.
Now we argue that $H^{(5)}$ should contain a bilinear in gluinos.
First, the original supercovariant $\hat{H}^{(5)}$ was derived only
for the IIb supergravity, in which no D-brane fields exist. It has
a term $-{1\over 16}\bar{\lambda}\gamma_{M_1\dots M_5}\lambda$. 
When 7-branes are present, $\lambda$ ceases to be massless, and
the new 8 dimensional dilatino must come as a linear combination
of gauginos \dl. On a general ground, not only one expects the presence
of bilinear of the new dilatino, one also expects bilinear of
all other gauginos. So we deduce that  $\hat{H}^{(5)}$ contains
a term $\tr\bar{\chi}\gamma_{\mu_1\dots \mu_5}\chi$, where indices
$\mu_i$ are tangent to 7-brane world-volume. This term is in perfect
harmony with the coupling \coupd, since supersymmetrization of \coupd\ 
eventually will force us to modify $H^{(5)}$ to include this bilinear.
We shall denote this bilinear simply by $\chi^{(5)}$.

On the world-volume of 7-branes, $\chi^{(5)}$ is dual to $\chi^{(3)}$.
If gluino condensation happens in such a way that $\chi^{(3)}$ has
a component only in the compact part of the world-volume, then 
coupling \coupd\ and $\delta\psi_\mu\ne 0$ indicate that SUSY is broken.
Obviously $\delta\psi_M\sim \gamma_M\hat{\gamma}_5$, where $\hat{\gamma}_5$
is the product of four gamma matrices $\gamma_\mu$, so the ``diagonal part''
$\gamma_M\psi_M$ must be part of the goldstino.

To understand this issue clearly, first we need to see how the massless
gluinos in 4 dimensions arise upon compactification. Apparently, the
gluinos are coupled to spin connection on the 4D sub-manifold $\Sigma$
which 7-branes  wrap around. The spin connection is the one induced from
that of the base space $B$ which has three complex dimensions. 
This can not be the whole story, as we have seen that generically 
$B$ is not Ricci flat so $\Sigma$ is not Ricci flat either. As a consequence
a gaugino can not have a zero mode if it is coupled only to the
spin connection. It must be that gauginos are topologically twisted,
as suggested in \bsv. The fact that $\Sigma$ is a singular locus of the 
elliptic fibration on $B$ implies that $\Sigma$ is a holomorphic cycle.
The usual criterion of supersymmetric cycle \bbs\ is satisfied. 
To see this, let us write down the condition for a four circle
\eqn\suc{\left(1 \pm {1\over 4!}h^{-1/2}\epsilon^{\alpha_1\dots\alpha_4}
\p_{\alpha_1}X^{m_1}\dots\p_{\alpha_4}X^{m_4}\gamma_{m_1\dots m_4}
\right)\epsilon=0,}
where we used $X^m$ as the real coordinates of $B$ and $y^\alpha$ as 
real coordinates of $\Sigma$, $h=\det(h_{\alpha\beta})$ and $h_{\alpha\beta}$
is the induced metric on $\Sigma$. Assuming that $B$ is K\"ahler,
so as a holomorphic sub-manifold $\Sigma$ is also K\"ahler.
Switch to complex coordinates, and use
the fact that $\epsilon$ is annihilated by ``holomorphic'' gamma
matrices $\gamma_i$, it is straightforward to see that the above
condition is satisfied for the minus sign.

Thus, the gaugino zero mode must be certain reduction of $\epsilon$
to a spinor on $\Sigma$. Let $\gamma_a$ $a=1,2$ denote the ``holomorphic''
gamma matrices on $\Sigma$, and let the reduction be $\psi$. Apparently,
for an elliptic fibration when $\gamma_i\epsilon=0$, 
then $\gamma_a\psi=0$. $\psi$ can be assigned with a definite chirality
on $\Sigma$. It is easy to see that the bilinear $\chi^{(3)}$ vanishes, 
if one plugs $\psi$ directly into it. This is not surprising, since
$\chi^{(3)}$ is defined for a pseudo-Majorana fermion in 8 dimensions.
Such a pseudo-Majorana spinor, being a zero mode of the twisted
Dirac operator, can in general give rise to a nonvanishing $\chi^{(3)}$
with components $\chi^+\gamma_{ab\bar{c}}\chi$ and 
$\chi^+\gamma_{a\bar{b}\bar{c}}\chi$. Dualizing against rank 4 antisymmetric
tensor in 4 Euclidean space, equivalently these components can be written
as two vectors $f^{\bar{a}}$ and $f^a$, and $f^{\bar{a}}$ is complex
conjugate of $f^a$.

Note that, although both in the F-theory context and in the heterotic
string, it takes bilinear $\chi^{(3)}$ to trigger SUSY breaking,
the details are quite different. In the heterotic string, $\chi^{(3)}$
has only purely holomorphic component and purely anti-holomorphic
component, both are covariantly constant as guaranteed by Calabi-Yau
threefold. In F-theory, $\chi^{(3)}$ lives on the four dimensional
locus $\Sigma$, so it must have components of mixed type. 

If $\chi^{(3)}$ is covariantly invariant, one may derive strong
constraints on geometry of $\Sigma$. Equivalently, one requires
$f^a$ as well as $f^{\bar{a}}$ be covariantly invariant. For a 
induced K\"ahler metric, it is easy to see that this implies that
$f^a$ is holomorphic, and $f^{\bar{a}}$ anti-holomorphic. Note
that the original quantity $\chi^{(5)}$ can be written directly
in terms of these vectors. For example, $\bar{\chi}\gamma_{\mu_1
\dots \mu_4 a}\chi\sim \epsilon_{\mu_1\dots\mu_4}g_{a\bar{b}}f^{\bar{b}}$. 
In the above discussion, we have ignored a factor proportional to 
condensate of the gluinos in 4D noncompact spacetime.

We have shown that indeed gluino condensation triggers SUSY breaking.
This must be the case, given that sometimes F-theory compactified
on a Calabi-Yau fourfold is dual to the heterotic string compactified
on a Calabi-Yau threefold. This happens when the base space $B$ is
rationally ruled, that is, it is a holomorphic fibration on another 
base space $B'$ with fiber $S^2$ \witten. Although technically 
the details are quite
different, we do expect discussions concerning SUSY breaking in
the heterotic string \drsw\ have parallel here, for example, the
cosmological constant remains vanishing. 

In the heterotic compactification, it is possible to consider
a nonvanishing torsion with a nontrivial warp factor \andy.
A similar situation was recently discussed in \bb\ for M-theory
compactified on an eight-manifold. We now show that compactification
with a warp factor is also possible for the IIb theory \foot{
We would like to thank Andy Strominger for bringing up the issue
of the warp factor.}. Here in addition to fibration function $\tau$ , 
a nonvanishing $H^{(5)}$ is necessary. 
A nonvanishing $H^{(5)}$ may be caused by the existence of D3-branes
transverse to the compact six-manifold. When some D3-branes
are introduced, their charge is assumed to be cancelled by that
caused by curved geometry, for instance the orbifold points.
Let $\hat{G}$ be the original
10D IIb metric, and let the compactification is such that
$\hat{G}=\exp (2\sigma )G$, where $\sigma$ is a function on the
base space $B$. The covariant derivative becomes
$$\hat{D}_M\epsilon =(D_M+{1\over 2}\gamma_M^{\quad N}
\p_N\sigma )\epsilon,$$
where the gamma matrices are defined against metric $G$. Of course 
$D_M$ contains the auxiliary $U(1)$ connection.
The SUSY transformation \gravi\ now becomes
\eqn\gravit{\delta\psi_M=D_M\epsilon +{1\over 2}\gamma_M^{\quad N}
\p_N\sigma\epsilon -{i\over 480}e^{-4\sigma}\gamma^{M_1\dots
M_5}\gamma_M\epsilon H^{(5)}_{M_1\dots M_5}.}

To preserve the 4D Minkowski spacetime, $D_\mu\epsilon=0$. Then
$\delta\psi_\mu=0$ demands a nonvanishing $H^{(5)}_{m_1\dots m_5}$
with indices $m_i$ tangent to $B$. Define
$$H^m={1\over 5!}g^{-1/2}\epsilon^{m_1\dots m_5 m}H^{(5)}_{m_1\dots
m_5},$$
\gravit\ can be re-written as
\eqn\graviti{\delta\psi_M=D_M\epsilon+{1\over 2}\gamma_M^{\quad
m}\p_m\sigma\epsilon -{1\over 2}e^{-4\sigma}H^m\gamma_m\hat{\gamma}_5
\gamma_M\epsilon,}
where we have used the self-duality of $H^{(5)}$, and 
$\hat{\gamma}_5=i\gamma^{0123}$. Taking $M=\mu$ in \graviti, we find
$$\gamma^m(\p_m\sigma -e^{-4\sigma}H_m\hat{\gamma}_5)\epsilon=0,$$
which is satisfied if
\eqn\prec{(\p_m\sigma -e^{-4\sigma}H_m\hat{\gamma}_5)\epsilon=0.}
Taking $M=m$ in \graviti\ and substituting \prec\ into it, one
finds
\eqn\ccon{\tilde{D}_m\epsilon=(D_m-2\p_m\sigma+{3\over 2}\gamma_m
\gamma^n\p_n\sigma) \epsilon =0.}
Again, in order to satisfy $\delta\lambda=0$, we assume 
that $B$ is a complex manifold and use $z^i$ as its complex 
coordinates, so that $\gamma_i\epsilon=0$. Written in term of
the complex coordinates, the covariant derivatives $\tilde{D}_m$
are
\eqn\covd{\eqalign{\tilde{D}_i&=D_i+\p_i\sigma,\cr
\tilde{D}_{\bar{i}}&=D_{\bar{i}}+{3\over 2}g^{i\bar{l}}\gamma_{\bar{j}\bar{l}}
\p_i\sigma-2\p_{\bar{i}}\sigma,}}
where we have used the fact that $\gamma_i\epsilon=0$.

The integrability condition $[\tilde{D}_i, \tilde{D}_j]
\epsilon=0$ yields $R_{ij}\epsilon=0$. And the integrability condition 
$[\tilde{D}_{\bar{i}}, \tilde{D}_{\bar{j}}]\epsilon=0$ yields
\eqn\fint{\left(R_{\bar{i}\bar{j}}+{3\over 2}(g^{i\bar{l}}\gamma_{
\bar{j}\bar{l}}D_{\bar{i}}\p_i\sigma -g^{i\bar{l}}\gamma_{
\bar{i}\bar{l}}D_{\bar{j}}\p_i\sigma)\right)\epsilon=0.}
An immediate consequence of this equation is that $B$ can not be
K\"ahler. If it were, then $R_{\bar{i}\bar{j}}=0$, and we would have
the condition, after carefully analyzing the above equation,
$$g^{\bar{i}l}D_{\bar{j}}\p_l\sigma =\delta^{\bar{i}}_{\bar{j}}
g^{l\bar{l}} D_{\bar{l}}\p_l\sigma,$$
which can not hold, as can be seen by taking the trace over $\bar{i},
\bar{j}$. So, in order to satisfy \fint, $R_{\bar{i}\bar{l}}^{\quad \bar{l}
\bar{k}}$ can not be vanishing. Finally, the integrability condition 
$[\tilde{D}_i, \tilde{D}_{\bar{j}}]\epsilon=0$ says
\eqn\sint{\left(R_{i\bar{j}}-F_{i\bar{j}}+{3\over 2}\p_i(g^{k\bar{l}}
\gamma_{\bar{j}\bar{l}}
\p_k\sigma)-3\p_i\p_{\bar{j}}\sigma\right)\epsilon =0.}
It should be possible to choose a metric and $\sigma$ to satisfy
both \fint\ and \sint. We shall leave a detailed analysis of these
equations, and possible gluino condensation on both 7-branes and
3-branes to the future.

Gluino condensation is a phenomenon of low energy field theory.
It would be interesting to explore interconnections between gluino
condensation discussed herein and genuinely stringy mechanism,
such as D-instanton generated superpotential discussed in \witten.
One might argue that the two are completely different phenomena.
However, as observed in the second reference of \sen\ and \nati,
closed string nonperturbative backgrounds can be probed by
field theory effects of D-branes, it is then possible that 
sometimes stringy effects and field theory effects are one and 
the same thing.

\noindent {\bf Acknowledgments} This work stemmed directly from the work 
\dl. We wish to thank Mike Douglas for discussions and on-going 
collaboration, and Andy Strominger for comments on a preliminary draft
which helped to improve this paper. This work was supported by DOE grant 
DE-FG02-91ER40688-Task A.

\listrefs

\end